\documentclass[12pt]{article}
\usepackage{pdproc}

\begin{document}

\setlength{\unitlength}{1mm}







\def \ni{\noindent}
\def \up{\bf}
\def \znbb {$0\nu\beta\beta$ }
\def \onbb {$0\nu\beta\beta$}
\def \tnbb {$2\nu\beta\beta$ }
\def \be {\begin{equation}}
\def \ee {\end{equation}}
\def \HM {HEIDEL\-BERG-MOSCOW~}
\def \GTF {GENIUS-Test-Facility~}
\def \etal {~et al.}

\renewcommand{\thefootnote}{\alph{footnote}}
 
\title{FIRST EVIDENCE FOR \\
	NEUTRINOLESS DOUBLE BETA DECAY -- \\
	AND WORLD STATUS OF DOUBLE BETA EXPERIMENTS}

\author{HANS VOLKER  KLAPDOR-KLEINGROTHAUS}

\address{Max-Planck-Institut f\"ur Kernphysik, PO 10 39 80,\\
  D-69029 Heidelberg, Germany, H.Klapdor@mpi-hd.mpg.de, 
	$http://www.mpi-hd.mpg.de.non\_acc/$}

\abstract{
        Nuclear double beta decay provides an extraordinarly 
        broad potential to search for beyond-standard-model physics. 
        {\it The occurrence of the neutrinoless decay} (\onbb) mode 
        has fundamental consequences: first {\bf total lepton number 
        is not conserved}, and second, 
        {\bf the neutrino is a Majorana particle}. 
        Further the effective mass measured allows to put an absolute 
        scale of the neutrino mass spectrum. 
        In addition, {\it double beta experiments yield  
        sharp restrictions also for other beyond standard model physics}. 
        These include SUSY models (R-parity breaking and conserving), 
        leptoquarks (leptoquark-Higgs coupling), 
        compositeness, left-right symmetric models 
        (right-handeld W boson mass), test of special relativity and 
        of the equivalence principle in the neutrino sector and others. 
        {\bf First evidence for neutrinoless double beta decay 
        was given in 2001, by the \HM experiment}. 
        The \HM experiment is the {\it by far most sensitive} \znbb 
        experiment since more than 10\,years. 
        It is operating 11\,kg of enriched $^{76}{Ge}$ 
        in the GRAN SASSO Underground Laboratory. 
        The analysis of the data 
        taken from 2 August 1990 - 20 May 2003, is presented here.  
        The collected statistics is 71.7\,kg\,y. 
        The background achieved in the energy region 
        of the Q value for double beta decay 
        is 0.11\,events/\,kg\,y\,keV.
        {\it The two-neutrino accompanied half-life} is determined 
        on the {\it basis of more than 100 000\,events} 
        to be $(1.74.^{+0.18}_{-0.16})\times 10^{21}~years.$ 
        {\bf The confidence level for the {\it neutrinoless} signal  
        is 4.2\,$\sigma$ level 
	(more than 5$\sigma$ in the pulse-shape-selected spectrum)}. 
	The half-life is 
        $T_{1/2}^{0\nu}=(1.19^{+0.37}_{-0.23})\times 10^{25}~years.$ 
        {\bf The effective neutrino mass deduced 
        is (0.2 - 0.6)\,eV (99.73\% c.l.)}, 
        with the consequence that neutrinos have degenerate masses, 
	and consequently still considerably, 
	and contribute to hot dark matter in the Universe. 
        The sharp boundaries for other beyond SM physics, 
        mentioned above,  
        are comfortably competitive to corresponding 
        results from high-energy accelerators like TEVATRON, HERA, etc. 
	Some {\it discussion} is given on future $\beta\beta$ experiments.}{}{} 

\normalsize\baselineskip=15pt

\section{Introduction}


        Since 40 years huge experimental efforts have gone 
        into the investigation of nuclear double beta decay 
        which probably is the most sensitive way to look for (total) 
        lepton number violation and probably the only way 
        to decide the Dirac or Majorana nature of the neutrino. 
        It has further perspectives 
        to probe also other types of beyond standard model physics. 
        This thorny way has been documented recently 
        in some detail 
\cite{KK60Y,Kla95/98,KK-SprTracts00}.

        With respect to half-lives to explore lying, with the order 
        of 10$^{25}$\,years, in a range  
        on 'half way' to that of proton decay, 
        the two main experimental problems were to achieve 
        a sufficient amount of double beta emitter material (source 
        strength) and to reduce the background in such experiment 
        to an extremely low level. 
        In both directions large progress has been 
        made over the decades. 
        While the first experiment using source as detector 
\cite{Goldh66}, 
        had only grams of material to its disposal 
        (10.6\,g of CaF$_2$), in the last years up 
        to more than 10\,kg of enriched emitter material 
        have been used. 
        Simultaneously the background of the experiments 
        has been reduced strongly over the last 40\,years. 
        For example, 
        compared to the  first Germanium $\beta\beta$ experiment 
\cite{1DBD-exp}, 
        working still with natural Germanium, containing 
        the double beta emitter $^{76}{Ge}$ only with 7.8\%, 
        40\,years later the background in the \HM experiment 
        is reduced by a factor of 10$^4$.

        The final dream behind all these efforts was less 
        to see a standard-model allowed second-order effect 
        of the weak interaction in the nucleus - the 
        two-neutrino-accompanied decay mode -  
        which has been observed meanwhile for about 10 nuclei - 
        (see e.g. 
\cite{KK60Y})
        but to observe neutrinoless double beta decay, and with
        this a first hint of beyond standard model physics, 
        yielding at the same time a solution 
        of the absolute scale of the neutrino mass spectrum. 

\section{Performance of the Experiment and Data Taking}

\subsection{General}

        The \HM experiment, proposed already in 1987 
\cite{Prop87-HM-HVKK}, 
        has been looking  
        for double beta decay of $^{76}{Ge}$ since August 1990 
        until November 30, 2003 in the Gran Sasso Underground Laboratory. 
        It was using the largest source strength of all double beta 
        experiments at present, and has reached a record
        low level of background, not only 
        for Germanium double beta decay search.
        It has 
        demonstrated this during more than a decade of measurements  
        and is since more then ten years the most sensitive
        double beta decay experiment worldwide. 
        The experiment was since 2001 operated only 
        by the Heidelberg group, which also performed 
        the analysis of the experiment from its very beginning. 

        The experiment has been carried out with five high-purity 
        p-type detectors 
        of Ge enriched to 86\% in the isotope $^{76}{Ge}$ 
        (in total 10.96 kg of active volume). 
        These were the first enriched high-purity Ge detectors ever produced. 
        So, the experiment starts from the cleanest thinkable source 
        of double beta emitter material, 
        which at the same time is used as detector of $\beta\beta$ events.

        A description of the experimental details 
        has been given in 
\cite{KK-NewAn-PL04,KK-NewAn-NIM04,Bey03-BB,HDM97}. 
        This will not be repeated in this paper, instead we concentrate 
        on the results and their consequences. 
        But let us just mention some of the most important features 
        of the experiment here.

        1. Since the sensitivity for the \znbb half-life is
        $T^{0\nu}_{1/2} \sim a  \times \epsilon \sqrt{\frac{Mt}{\Delta EB}}$ 
         (and 
        $\frac{1}{\sqrt{T^{0\nu}}} \sim \langle m_\nu \rangle$), 
        with $a$ denoting the degree of enrichment, $\epsilon$ 
        the efficiency of the detector for detection of a double beta event, 
        $M$ the detector (source) mass, $\Delta E$ the energy resolution, 
        $B$ the background and $t$ the measuring time, 
        the sensitivity of our 11\,kg {\it of enriched} $^{76}{Ge}$ 
        experiment corresponds to that of an at least 1.2\,ton 
        {\it natural} Ge experiment. After enrichment -  
        the other most important parameters of a $\beta\beta$ 
        experiment are: energy resolution, 
        background and source strength.

        2. The high energy resolution of the Ge detectors of 0.2$\%$ 
        or better, assures that there 
        is no background for a \znbb line from the two-neutrino 
        double beta decay in this experiment, in contrast to most 
        other present experimental approaches, 
        where limited energy resolution is a severe drawback.

        3. The efficiency of Ge detectors for detection 
        of \znbb decay events is close to 100\,$\%$ (95\%, see 
\cite{KK-NewAn-NIM04}).

        4. The source strength in this experiment of 11\,kg 
        is the largest source strength ever operated in 
        a double beta decay experiment.

        5. The background reached in this experiment, is  
        0.113$\pm$0.007 events /kg\,y\,keV (in the period 1995-2003)  
        in the \znbb decay region 
        (around Q$_{\beta\beta}$). 
        This is the lowest limit ever obtained 
        in such type of experiment.

        6. The statistics collected in this experiment during 
        13 years of stable running is the largest ever collected 
        in a double beta decay experiment. 
        The experiment took data during $\sim$ 80\% of its installation time.
        
        7. The Q value for neutrinoless double beta decay has been 
        determined recently with  
        high precision 
\cite{New-Q-2001}.

\section{Data and Analysis}

\begin{figure}[ht] %
\begin{picture}(100,130) 
\put(40,0){\includegraphics{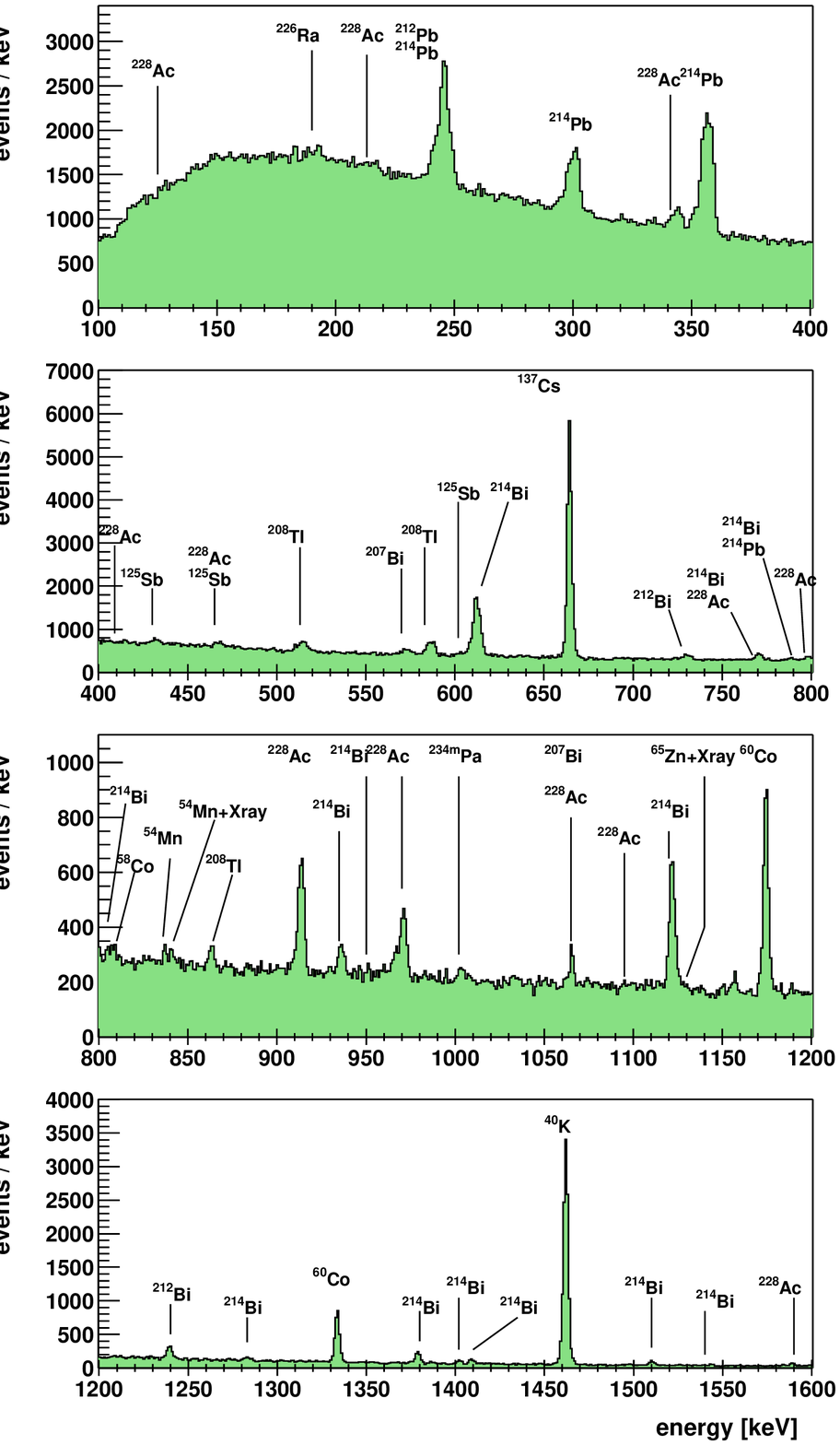}}
\end{picture}
\caption{The total sum spectrum measured over the full energy range 
        (low-energy part) of all five detectors (in total 10.96\,kg enriched 
        in $^{76}{Ge}$  to 86\%) - for the period 
        November 1995 to May 2003.}
\label{fig:LowAll90-03}
\end{figure}

\begin{figure}[ht] 
\begin{picture}(100,130) 
\put(40,0){\includegraphics{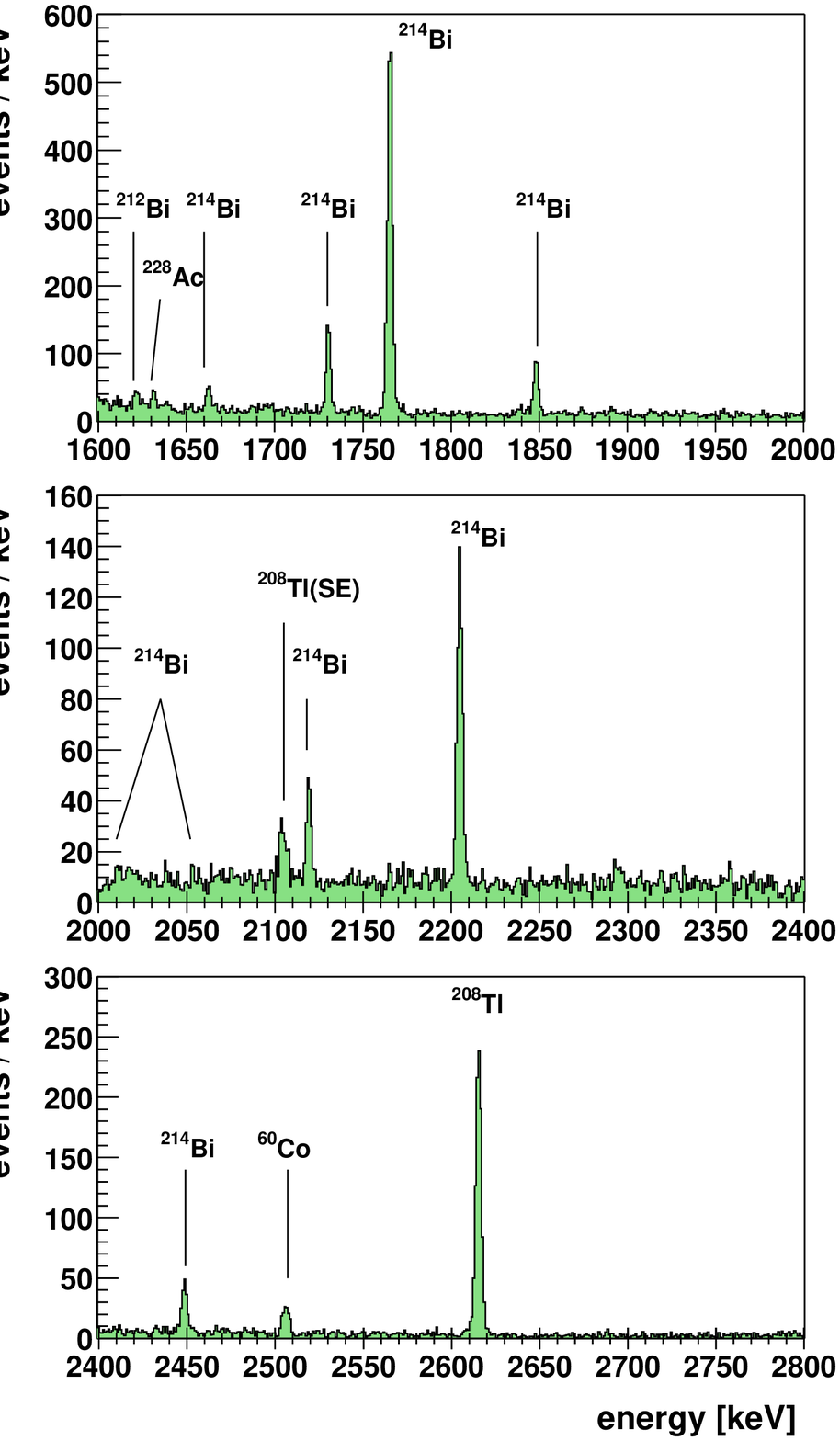}}
\end{picture}
\caption{The total sum spectrum measured over the full energy range 
        (higher energy part) of all five detectors 
        (in total 10.96\,kg enriched 
        in $^{76}{Ge}$  to 86\%) - for the period 
        November 1995 to May 2003.}
\label{fig:HighAll90-03}
\end{figure}

        Figs.%
\ref{fig:LowAll90-03},\ref{fig:HighAll90-03} 
        show the total sum spectrum measured over the full energy range 
        of all five detectors for the period 
        November 1995 to May 2003. 
        The identified lines are indicated with their source 
        of origin (for details see%
\cite{KK-Doer03}). 

\begin{figure}[ht] 
\begin{picture}(100,85) 
\put(-5,0){\includegraphics{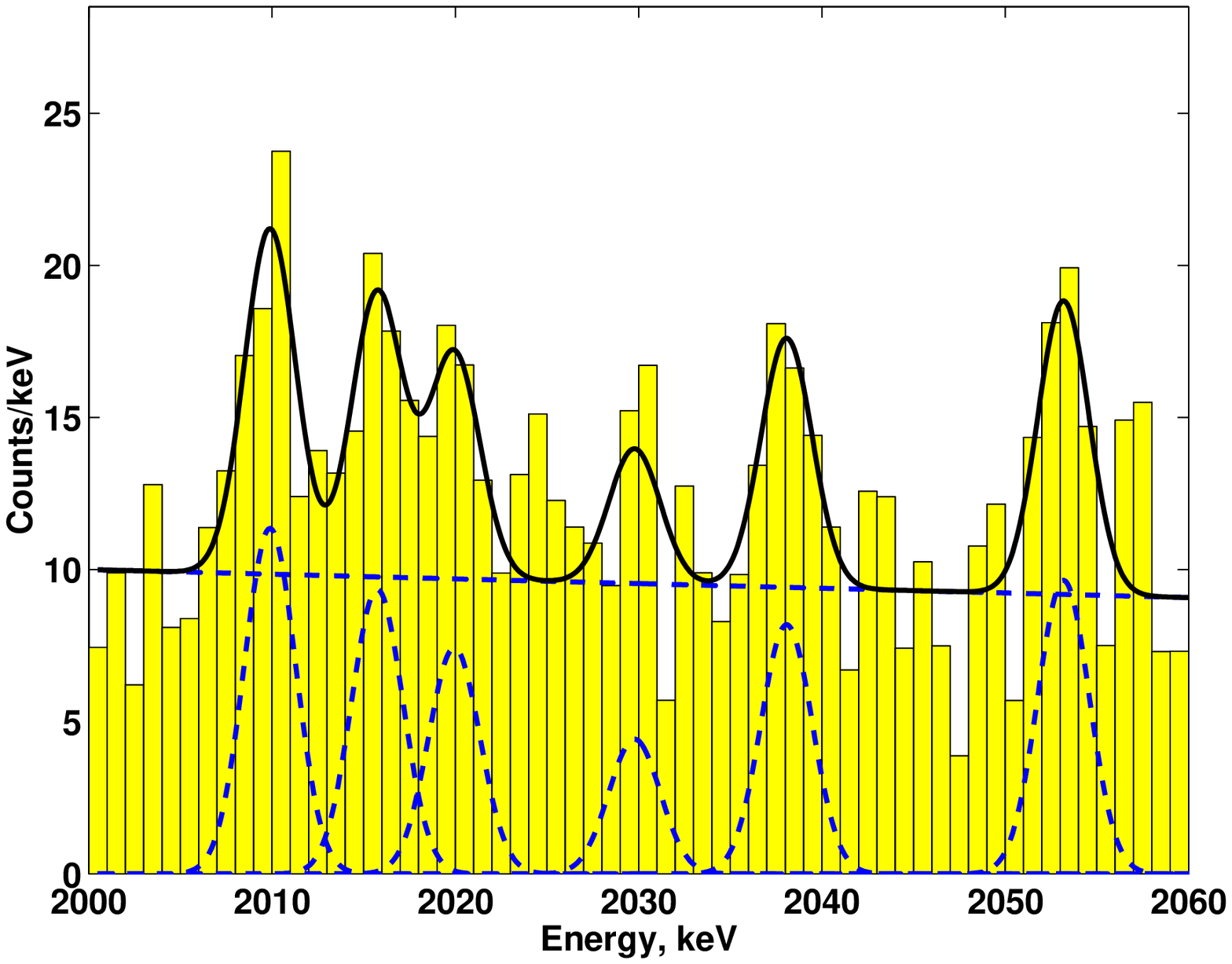}}
\put(71,0){\includegraphics{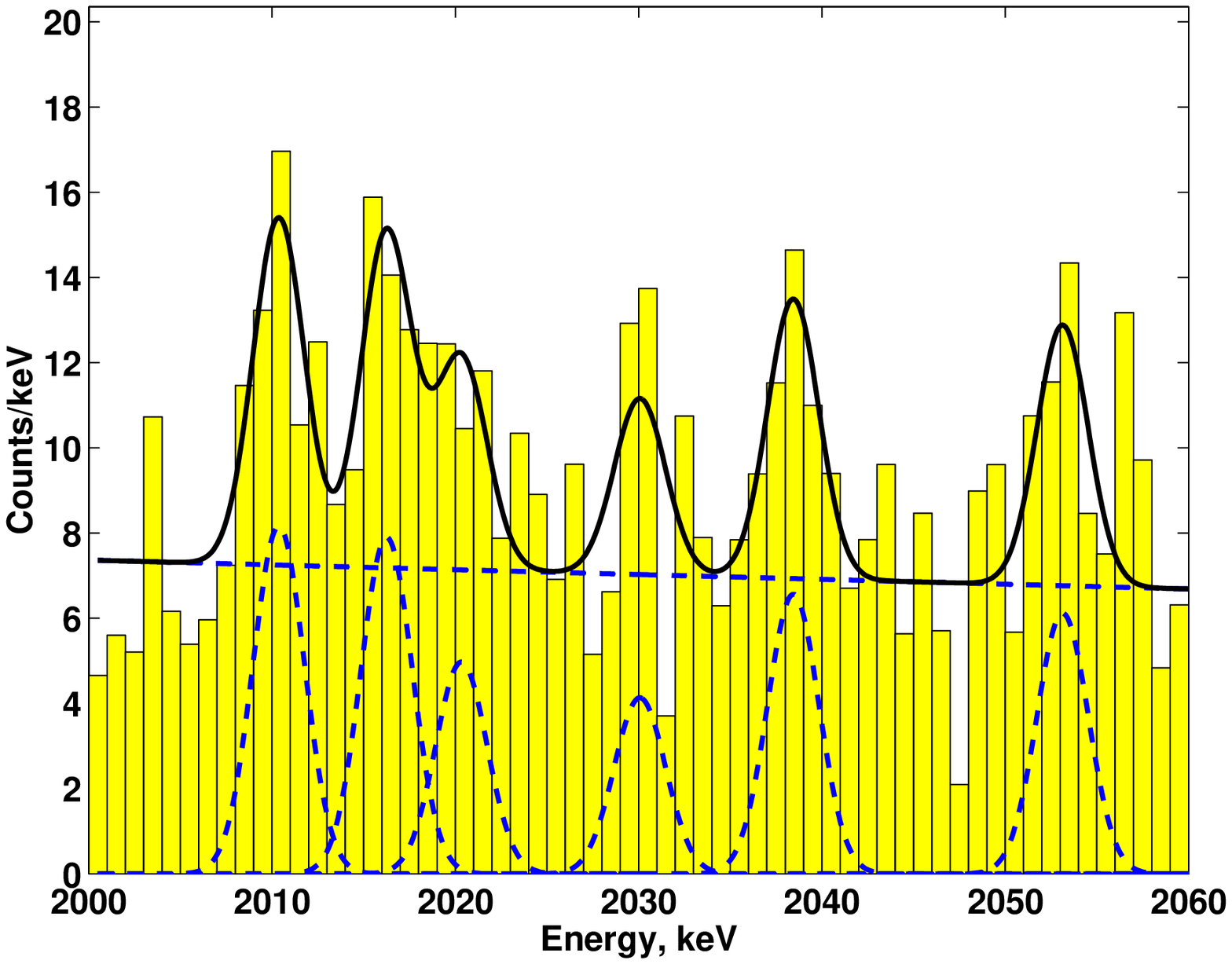}}
\end{picture}

\vspace{-2.5cm}
\caption{
        The total sum spectrum of all five detectors 
        (in total 10.96\,kg enriched in $^{76}{Ge}$), for the period: 
        \underline{left:} November 1990 to May 2003 (71.7\,kg\,y) 
        in the range 2000 - 2060\,keV. \protect\newline  
        \underline{right:} - November 1995 to May 2003 (56.66\,kg\,y) 
        in the range 2000 - 2060\,keV and its fit (see section 3.2).}
\label{fig:Sum90-03}
\end{figure}

\begin{figure}[ht!] 
\begin{picture}(100,100) 
\put(30,20){\includegraphics{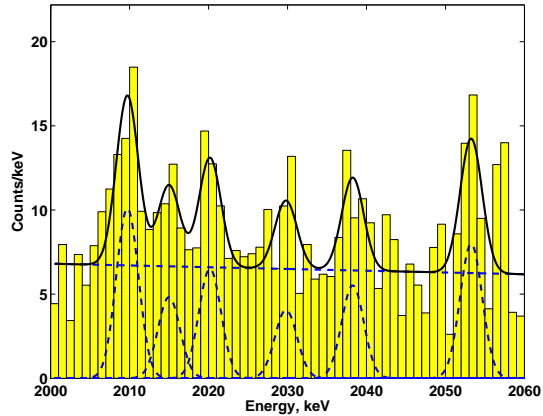}}
\end{picture}

\vspace{-4.5cm}
\caption{The total sum spectrum of all five detectors 
        (in total 10.96\,kg enriched in $^{76}{Ge}$), 
        in the range (2000 - 2060)\,keV 
        and its fit, for the period 
        August 1990 to May 2000 (50.57\,kg\,y). 
}
\label{fig:Sum90-00}
\end{figure}

        Figs.%
\ref{fig:Sum90-03},\ref{fig:Sum90-00}  
        show the part of the spectrum around Q$_{\beta\beta}$, 
        in the range 2000 - 2060\,keV, measured 
        in the period August 1990 to May 2003 
        and November 1995 to May 2003. 
        Non-integer numbers in the sum spectra are simply a binning effect.


                \subsection{Energy Calibration}

        Precise energy calibration for all detectors 
        before summing the individual 2142 runs 
        taken with the detectors, 
        and finally summing the sum spectra of the different detectors 
        (in total summing 9 570 data sets)   
        is decisive to achieve a good
        energy resolution of the total spectrum, 
        and an optimum sensitivity of the experiment. 
        For details see 
\cite{KK-NewAn-PL04,KK-NewAn-NIM04,Bey03-BB}. 
        A list of the energies of the identified lines (Figs.%
\ref{fig:LowAll90-03},\ref{fig:HighAll90-03}) 
        is given in a recent paper 
\cite{KK-Doer03}, 
        here we concentrate on the range 
        of interest around Q$_{\beta\beta}$.


\subsection{Analysis of the Spectra}

        In the measured spectra (Figs.%
\ref{fig:Sum90-03}$\div$\ref{fig:Sum90-00})
        we see in the range around $Q_{\beta\beta}$ the
        $^{214}{Bi}$ lines 
        at 2010.7, 2016.7, 2021.8, 2052.9\,keV,
        the line at $Q_{\beta\beta}$ and a candidate of a line 
        at $\sim$ 2030\,keV (see also
\cite{Bi-KK03-NIM,PL03})\footnote{The objections raised after our first paper 
\cite{KK02}
        concerning these lines and other points, by Aalseth et al. 
        (Mod.Phys.Lett.A17:1475-1478,2002 and hep-ex/0202018 v.1), 
        have been shown to be wrong already in 
\cite{KK-Repl-Crit26Kommis}
        and in 
\cite{KK02-Found},
        and later in 
\cite{Bi-KK03-NIM} and 
\cite{PL03}. 
        So this 'criticism' was already history, before we reached 
        the higher statistics presented in this paper.} 
        The spectra have been analyzed by {\it different methods}: 
        Least Squares Method, Maximum Likelihood Method (MLM) 
        and Feldman-Cousins Method. 
        The analysis is performed 
        {\it without subtraction of any background}. 
        We always process background-plus-signal 
        data since the difference between two Poissonian 
        variables does {\it not} produce a Poissonian
        distribution 
\cite{NIM99}. 
        This point has to be stressed, 
        since it is sometimes overlooked. 
        So, e.g., in 
\cite{Zdes} 
        a formula is developed making use 
        of such subtraction and as a consequence 
        the analysis given in 
\cite{Zdes} 
        provides overestimated standard errors.

        The large improvement of the present analysis (for details see 
\cite{KK-NewAn-PL04,KK-NewAn-NIM04,Bey03-BB}) 
        compared to our paper from 2001 
\cite{KK02,KK02-PN,KK02-Found},
        is clearly seen from Fig.  
\ref{fig:Sum90-00} 
        showing the {\it new} analysis 
        of the data 1990-2000, 
        as performed here -- 
        to be compared to the corresponding figure in 
\cite{KK02,KK02-PN,KK02-Found}.
        One reason lies in the stricter conditions for accepting data 
        into the analysis. 
        The spectrum in Fig.  
\ref{fig:Sum90-00}
        now corresponds to 50.57\,kg\,y 
        to be compared to 54.98\,kg\,y in 
\cite{KK02},
        for the same measuring period.
        The second reason is a better energy calibration 
        of the individual runs. The third reason 
        is the refined summing procedure of the individual data sets 
        mentioned above and the correspondingly better energy resolution 
        of the final spectrum. 
        (For more details see 
\cite{KK-NewAn-PL04,KK-NewAn-NIM04,Bey03-BB}). 
        The signal strength seen in the {\it individual} detectors 
        in the period 1990-2003 is shown in 
\cite{Bey03-BB}.

        We tested the confidence intervals calculated 
        by the fitting programs with numerical simulations (see 
\cite{KK-NewAn-PL04,KK-NewAn-NIM04,Bey03-BB}). 
        As done earlier for other statistical methods 
\cite{KK02-PN,KK02-Found}, 
        we have simulated 100 000 spectra with Poisson-distributed 
        background and a Gaussian-shaped (Poisson-distributed) 
        line of given intensity, and have investigated, 
        in how many cases we find in the analysis the known intensities 
        inside the given confidence range. 
        The result 
        shows that the confidence levels determined 
        are correct within small errors (for details see 
\cite{KK-NewAn-NIM04,Bey03-BB}). 

\section{Results}

\subsection{Full Spectra}

\begin{figure}[ht] 
\begin{picture}(100,65) 
\put(30,0){\includegraphics{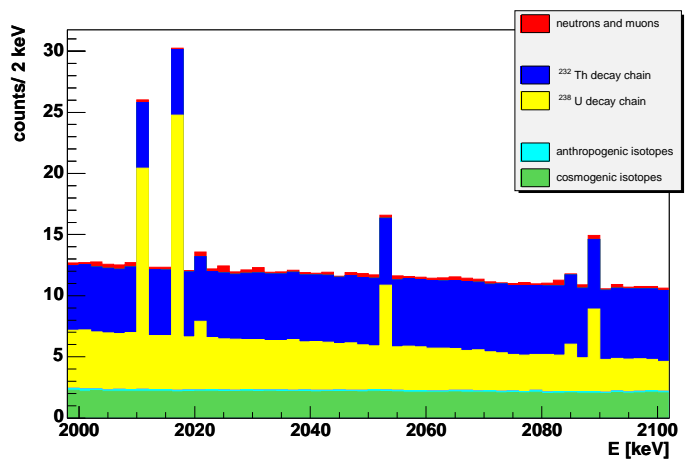}}
\end{picture}
\caption{\label{bb_under}
        Monte Carlo simulation of the background in the range of 
        Q$_{\beta\beta}$ by GEANT4, including all known sources 
        of background in the detectors and the setup. 
        This simulation 
\protect\cite{KK-Doer03} 
        seems to be the by far most extensive 
        and complete one ever made for any double beta experiment.
        The background around Q$_{\beta\beta}$ is expected 
        to be flat, the only lines visible should 
        be some weak $^{214}$Bi lines (from 
\protect\cite{KK-Doer03}).}
\end{figure}

        Figs. 
\ref{fig:Sum90-03}$\div$\ref{fig:Sum90-00}
        show together with the measured spectra 
        in the range around Q$_{\beta\beta}$ (2000 - 2060\,keV), 
        the fit by the least-squares method.
        A linear decreasing shape of the background as function 
        of energy was chosen corresponding to the complete simulation 
        of the background performed in  
\cite{KK-Doer03} 
         by GEANT4 (see Fig. 
\ref{bb_under}).


        In the fits in Figs. 
\ref{fig:Sum90-03},\ref{fig:Sum90-00}
        the peak positions, widths and intensities 
        are determined simultaneously, and also the {\it absolute} level 
        of the background. 

        The signal at Q$_{\beta\beta}$ 
        in the full spectrum at $\sim$ 2039\,keV 
        reaches a 4.2 $\sigma$ confidence level 
        for the period 1990-2003 (28.8 $\pm$ 6.9)\,events, 
        and of 4.1 $\sigma$ 
        for the period 1995-2003 (23.0 $\pm$5.7)\,events. 
        The results of the new analysis are consistent 
        with the results given in 
\cite{KK02,KK02-PN,KK02-Found}.
        The intensities of all other lines are given in 
\cite{KK-NewAn-NIM04,Bey03-BB}.

        We have given a detailed comparison 
        of the spectrum measured in this experiment 
        with other Ge experiments in 
\cite{PL03}.
        It is found that the most sensitive experiment 
        with natural Ge detectors 
\cite{Caldw91}, 
        and the first experiment using enriched (not yet high-purity) 
        $^{76}{Ge}$ detectors 
\cite{vasenko}
        find essentially the same background lines 
        ($^{214}{Bi}$ etc.), but {\it no} indication 
        for the line near Q$_{\beta\beta}$. 
        This is consistent with the rates expected 
        from the present experiment due 
        to their lower sensitivity: 
        $\sim$ 0.7 and $\sim$ 1.1 events, respectively. 
        It is also consistent with the result 
        of the IGEX $^{76}{Ge}$ experiment 
\cite{igex2}, 
        which collected only a statistics of 8.8\,kg\,y, 
        before finishing in 1999, and which should expect $\sim$ 2.6\,events, 
        which they might have missed. 
        Their published half-life limit is overestimated as result 
        of an arithmetic mistake (see 
\cite{KK-PhysRC}).

\begin{figure}[ht!] 
\begin{picture}(100,110) 
\put(30,0){\includegraphics{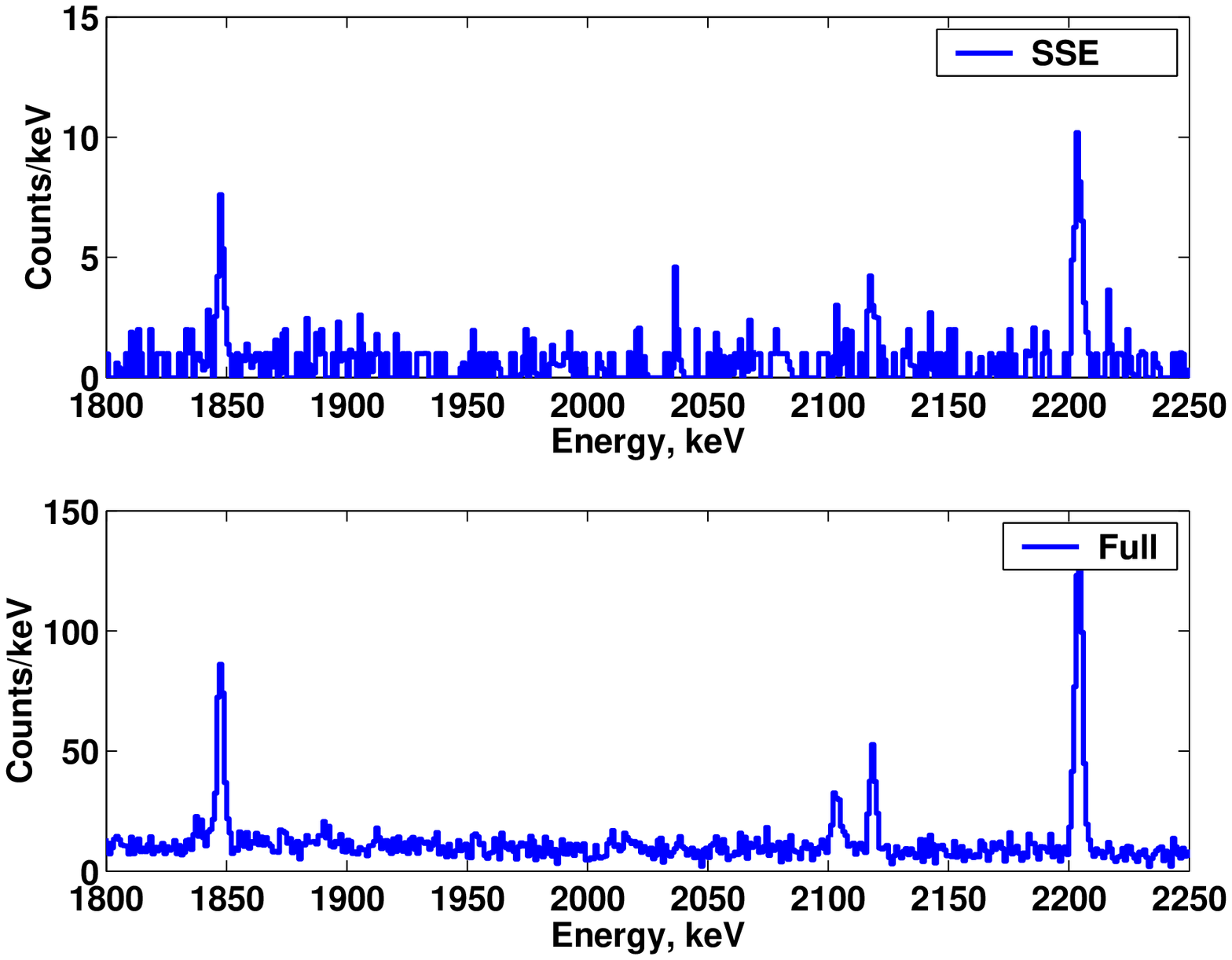}}
\put(38,-20){\includegraphics{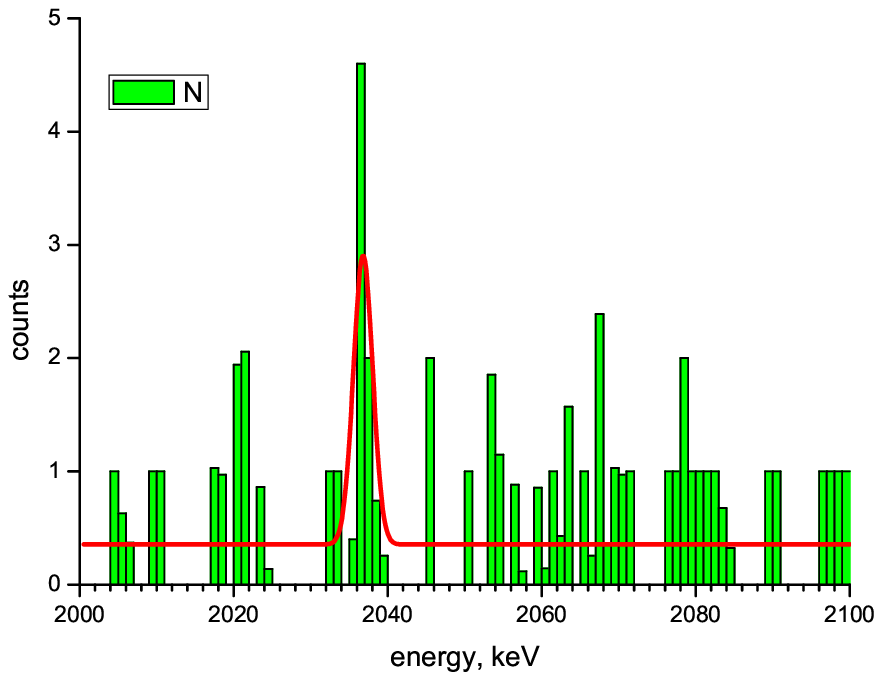}}
\end{picture}

\vspace{2.cm}
\caption{Top, upper part: 
        The pulse-shape selected spectrum of single site events 
        measured with detectors 2,3,4,5 from 1995-2003, see text. 
        Top, lower part: The full spectrum measured with detectors 2,3,4,5 
        from 1995-2003. 
        Bottom: As in top figure, upper part, but energy range 2000-2100\,keV.}
\label{fig:Full-NN-95-03}
\end{figure}

\subsection{Time Structure of Events}

        There are at present {\it no other} running experiments 
        (with reasonable energy resolution) 
        which can - not to speak about their lower sensitivity - 
        {\it in principle} give {\it any further-going} information 
        in the search for double beta decay than shown up to this point: 
        namely a line at the correct energy Q$_{\beta\beta}$. 
        Also most future projects cannot determine more. 
        The \HM experiment developed some 
        {\it additional tool} of independent verification. 
        The method is to exploit the time structure of the events 
        and to select $\beta\beta$ events by their pulse shape.
        The result is shown in Fig.%
\ref{fig:Full-NN-95-03}.

\begin{figure}[ht] 
\begin{picture}(100,105) 
\put(25,0){\includegraphics{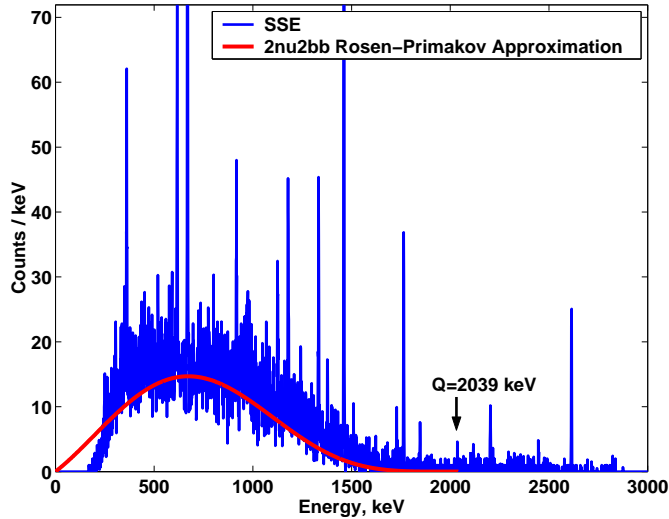}}
\end{picture}

\vspace{-3.cm}
\caption{The pulse-shape selected spectrum 
        measured 
        with detectors 2,3,4,5 from 1995$\div$2003 
        in the energy range of (100$\div$3000)\,keV, see text. 
}
\label{fig:beta-beta}
\end{figure}

\begin{table}[ph!]
\caption{Half-life for the neutrinoless decay mode 
        and deduced effective neutrino mass 
        from the HEIDELBERG-MOSCOW experiment 
        (the nuclear matrix element of 
\protect\cite{Sta90} 
        is used). 
        Shown are in addition to various accumulated total 
        measuring times also the results for four 
        {\it non-overlapping} data sets: 
        the time periods 11.1995-09.1999 and 09.1999$\div$05.2003 for 
        {\it all} detectors, and the time period 1995$\div$2003 
        for two sets of detectors: 1+2+4, and 3+5.
        *) denotes best value.
}
\begin{center}
{\footnotesize
\newcommand{\m}{\hphantom{$-$}}
\renewcommand{\arraystretch}{1.5}
\setlength\tabcolsep{3.5pt}
\begin{tabular}{|c|c|c|c|c|}
\hline
Significan-     &       Detectors               
&       ${\rm T}_{1/2}^{0\nu}   ~~[y]$  
& $\langle m \rangle $ [eV]     &       Conf. \\
        ce $[ kg\,y ]$  &       &       (3$\sigma$ range) 
&               (3$\sigma$ range)& level ($\sigma$)\\
\hline
\multicolumn{5}{|c|}{\bf Period 8.1990 $\div$ 5.2003}\\
\hline
        {\bf 71.7}    
&       1,2,3,4,5       
&       ${\bf (0.69 - 4.18) \times 10^{25}}$  
& {\bf (0.24 - 0.58)} 
&       {\bf 4.2}     \\
                &       
&       ${\bf 1.19 \times 10^{25}}$$^*$       
& {\bf 0.44$^*$}      
&       \\
\hline
\multicolumn{5}{|c|}{\it Period 11.1995 $\div$ 5.2003}\\
\hline
        56.66   
&       1,2,3,4,5
&       $(0.67 - 4.45) \times 10^{25}$  
& (0.23 - 0.59)
& 4.1   \\
 &      &       $1.17 \times 10^{25}$$^*$
& 0.45$^*$ 
&               \\
\hline
        51.39   
&       2,3,4,5
&       $(0.68 - 7.3) \times 10^{25}$   
& (0.18 - 0.58)
& 3.6   \\
 &      &       $1.25 \times 10^{25}$$^*$       
& 0.43$^*$ 
&               \\
\hline
        42.69   
&       2,3,5
&       $(0.88 - 4.84) \times 10^{25}$  
& (0.22 - 0.51)
& 2.9   \\
        &       
&       (2$\sigma$ range)       & (2$\sigma$ range)     &       \\
 &      &       $1.5 \times 10^{25}$$^*$        
& 0.39$^*$ 
&               \\
\hline
        28.27   
&       1,2,4
&       $(0.67 - 6.56) \times 10^{25}$  
& (0.19 - 0.59)
& 2.5   \\
        &       
&       (2$\sigma$ range)       & (2$\sigma$ range)     &       \\
 &      &       $1.22 \times 10^{25}$$^*$       
& 0.44$^*$  
&               \\
\hline
        28.39   
&       3,5
&       $(0.59 - 4.29) \times 10^{25}$  
& (0.23 - 0.63)
& 2.6   \\
        &       
&       (2$\sigma$ range)       & (2$\sigma$ range)     &       \\
 &      &       $1.03 \times 10^{25}$$^*$       
& 0.48$^*$  
&               \\
\hline
\multicolumn{5}{|c|}{\it Period 11.1995 $\div$ 09.1999}\\
\hline
        26.59   
&       1,2,3,4,5
&       $(0.43 - 12.28) \times 10^{25}$ 
& (0.14 - 0.73)
& 3.2   \\
 &      &       $0.84 \times 10^{25}$$^*$
& 0.53$^*$ 
&               \\
\hline
\multicolumn{5}{|c|}{\it Period 09.1999 $\div$ 05.2003}\\
\hline
        30.0    
&       1,2,3,4,5
&       $(0.60 - 8.4) \times 10^{25}$   
& (0.17 - 0.63)
& 3.5   \\
 &      &       $1.12 \times 10^{25}$$^*$
& 0.46$^*$ 
&               \\
\hline
\end{tabular}\label{Results1} }
\end{center}
\end{table}


        Here a {\it subclass} of shapes selected 
        by the neuronal net method used earlier 
\cite{KK02-Found,KK02-DARK02} 
        is shown. 
        Except a line which sticks out sharply near Q$_{\beta\beta}$, 
        {\it all} other lines are very strongly suppressed. 
        Fig.  
\ref{fig:Full-NN-95-03}
        also shows the full spectrum in this range. 
	{\bf When taking the range 2000-2100\,keV and 
	conservatively assuming all structures except 
	the line at Q$_{\beta\beta}$ 
	to be part of a constant background, a corresponding 
	fit yields a signal at Q$_{\beta\beta}$ 
	of more than 5$\sigma$ (see fig. 
\ref{fig:Full-NN-95-03})}.
        The method seems also to fulfill the criterium to select 
        properly the {\it continuous \tnbb spectrum} 
        (see Fig. 
\ref{fig:beta-beta}). 

        The energy of this line  
        determined by the spectroscopy ADC is  
        slightly below Q$_{\beta\beta}$, 
        but still within the statistical variation for a weak line (see 
\cite{PL03}). 
        This can be understood as result of ballistic effects 
        (for details see%
\cite{KK-SSE04}).
        The 2039\,keV line as a single site events signal 
        cannot be the double escape line of a $\gamma$-line 
        whose full energy peak would be expected at 3061\,keV,  
        since no indication of a line is found there in the spectrum 
        measured up to 8\,MeV (see  
\cite{KK-NewAn-NIM04,Bey03-BB}).

\section{Half-Life of Neutrinoless Double Beta Decay of~ $^{76}{Ge}$}

        We have shown in chapter 4 that 
        the signal found at Q$_{\beta\beta}$ is  
        consisting of single site events 
        and is not a $\gamma$ line.
        The signal does not occur 
        in the Ge experiments {\it not} enriched 
        in the double beta emitter $^{76}{Ge}$ 
\cite{Caldw91,Backgr-KK03-NIM,PL03},
        while neighbouring background lines appear consistently 
        in these experiments.

        On this basis we translate the observed numbers 
        of events into half-lives for neutrinoless double beta decay. 
        In Table
\ref{Results1} 
        we give the half-lives deduced from 
        the full data sets taken in the years 1995-2003 
        and in 1990-2003  
        and of some partial data sets.   
        In all cases the signal is seen consistently. 
        Also given are the deduced effective neutrino masses.

        The result obtained {\bf is consistent} with the limits 
        given earlier 
\cite{HDM01},
        and with the results given in 
\cite{KK02,KK02-PN,KK02-Found}.

        {\it Concluding} {\bf we confirm}, 
        with {\bf 4.2$\sigma$ (99.9973$\%$ c.l.) probability  
	(more than 5$\sigma$ in the pulse-shape selected spectrum)}, 
        our claim from 2001 
\cite{KK02,KK02-PN,KK02-Found} 
        {\bf of first evidence 
        for the neutrinoless double beta decay mode}.

\section{Consequences for Particle Physics, Neutrino Physics and\\ Other Beyond Standard Model Physics}


        {\bf Lepton number violation:}
        {\it The most important consequence} of the observation 
        of neutrinoless double beta decay is, that   
        {\bf lepton number is not conserved}. 
        This is fundamental for particle physics.\\ 

        {\bf Majorana nature of neutrino:} 
        Another fundamental consequence is that 
        {\bf the neutrino is a Majorana particle} (see, e.g. 
\cite{Sch82a,H_KK_Kov97-98}, but also \cite{Pilaft-KK04}).
        Both of these conclusions are {\it independent of any} 
        discussion of nuclear matrix elements.\\

        {\bf Effective neutrino mass}:
        The matrix element enters when we derive 
        a {\it value} for the effective neutrino mass 
        - making the {\it most natural assumption} 
        that the \znbb decay amplitude 
        is dominated by exchange of a massive Majorana neutrino. 
        The half-life for the neutrinoless decay mode 
        is under this assumption given by 
\cite{Sta90,Mut89}

$[T^{0\nu}_{1/2}(0^+_i \rightarrow 0^+_f)]^{-1}= C_{mm} 
\frac{\langle m \rangle^2}{m_{e}^2}
+C_{\eta\eta} \langle \eta \rangle^2 + C_{\lambda\lambda} 
\langle \lambda \rangle^2 +C_{m\eta} \langle \eta \rangle \frac{\langle m \rangle}{m_e}$ 

\centerline{
$+ C_{m\lambda}
\langle \lambda \rangle \frac{\langle m \rangle}{m_e}+C_{\eta\lambda}
\langle \eta \rangle \langle \lambda \rangle,$
}
\begin{equation}
\langle m \rangle = 
|m^{(1)}_{ee}| + e^{i\phi_{2}} |m_{ee}^{(2)}|
+  e^{i\phi_{3}} |m_{ee}^{(3)}|,~ 
\end{equation}
        where $m_{ee}^{(i)}\equiv |m_{ee}^{(i)}| \exp{(i \phi_i)}$ 
        ($i = 1, 2, 3$)  are  the contributions 
        to the effective mass $\langle m \rangle$
        from individual mass eigenstates, 
        with  $\phi_i$ denoting relative Majorana phases connected 
        with CP violation, and
        $C_{mm},C_{\eta\eta}, ...$ denote nuclear matrix elements squared, 
        which can be calculated, (see, e.g. 
\cite{KK60Y,Gro89/90,Mut88},  
        for a review).
        Ignoring contributions from right-handed weak currents, on the 
        right-hand side of eq.(1) only the first term remains.

        Using the nuclear matrix element from 
\cite{Sta90,Mut89}, 
        we conclude from the half-life  
        given above the effective mass 
        $\langle m \rangle $ 
        to be $\langle m \rangle $ 
        = (0.2 $\div$ 0.6)\,eV (99.73$\%$ c.l.), 
        with {\bf best value of $\sim$ 0.4\,eV}.

        The matrix element given by 
\cite{Sta90}
        was the {\it prediction closest to} the {\it later} measured \tnbb
        decay half-life 
        of $({1.74} ^{+0.18}_{-0.16})\times 10^{25}$\,y 
\cite{KK-Doer03,HDM97}. 
        It underestimates the 2$\nu$ matrix elements by 32\% 
        and 
        thus these calculations will also underestimate (to a smaller extent) 
        the matrix element for \znbb decay, 
        and consequently correspondingly overestimate 
        the (effective) neutrino mass. 
	The value for the effective mass thus in reality 
	will be somewhat lower, than deduced above, 
	down to {\bf $\sim$ 0.3 eV}. 
        Allowing conservatively for an uncertainty of the nuclear 
        matrix element of $\pm$ 50$\%$
        the range for the effective mass may widen 
        to $\langle m \rangle $ = (0.1 - 0.9)\,eV (99.73\% c.l.).


\begin{figure}[ht] 
\begin{picture}(100,80) 
\put(15,0){\includegraphics{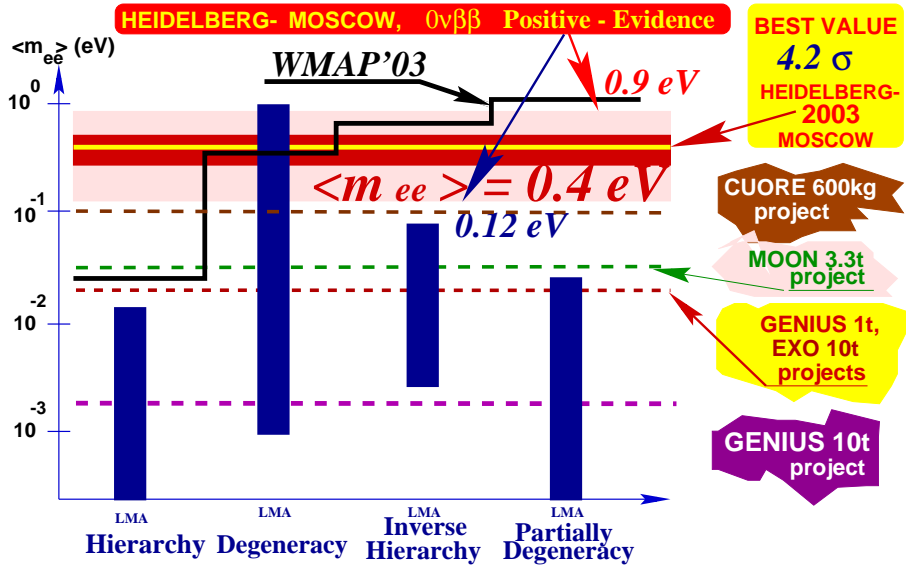}}
\end{picture}
\caption{The impact of the evidence obtained for neutrinoless 
        double beta decay (best value 
        of the effective neutrino mass 
        $\langle m \rangle$ = 0.4\,eV, 3$\sigma$ 
        confidence range (0.1 - 0.9)\,eV - 
        allowing already for an uncertainty of the nuclear 
        matrix element of a factor of $\pm$ 50$\%$) 
        on possible neutrino mass schemes. 
        The bars denote allowed ranges of $\langle m \rangle$ 
        in different neutrino mass scenarios, 
        still allowed by neutrino oscillation experiments (see 
\protect\cite{KK-Sark01,KK-Sark-WMAP03}). 
        All models except the degenerate one are excluded by the 
        new \znbb ~~decay result. 
        Also shown is the exclusion line from WMAP, 
        plotted for $\sum m_{\nu} < 1.0\, eV$ 
\protect\cite{Hannes03} 
        (which is according to 
\protect\cite{Teg03} 
        too strict). 
        WMAP does not rule out any of the neutrino mass schemes. 
        Further shown are the expected sensitivities 
        for the future potential double beta experiments 
        CUORE, MOON, EXO  
        and the 1 ton and 10 ton project of GENIUS 
\protect\cite{KK60Y,KK-SprTracts00,GEN-prop} 
        (from 
\protect\cite{KK-Sark-WMAP03}).
\label{fig:Jahr00-Sum-difSchemNeutr}}
\end{figure}


        {\bf Neutrinos degenerate in mass:}
        With the value deduced for the effective neutrino mass,  
        the HEIDELBERG-MOSCOW experiment excludes several 
        of the neutrino mass scenarios 
        allowed from present neutrino oscillation experiments
        (see Fig. 
\ref{fig:Jahr00-Sum-difSchemNeutr}, 
        and Fig.1 in 
\cite{KK-Sark-WMAP03}), 
        -- allowing only for degenerate   
        mass scenarios 
\cite{KK-Sark01,KK-Sark-WMAP03,KKPS}. 
        Degenerate mass scenarios had been discussed already earlier 
        (see e.g. 
\cite{Mina97,Yasuda99}).
        In connection with the L/E flatness of the electron-like 
        event ratio observed in Superkamiokande, degeneracy 
        has been discussed by 
\cite{Ahluw99}.

	{\bf Neutrinos as hot dark matter:} 
	The effective neutrino mass determined by \znbb decay allows 
	a considerable fraction of hot dark matter in the Universe 
	carried by neutrinos.

\begin{figure}[ht] 
\begin{picture}(100,70) 
\put(5,0){\includegraphics{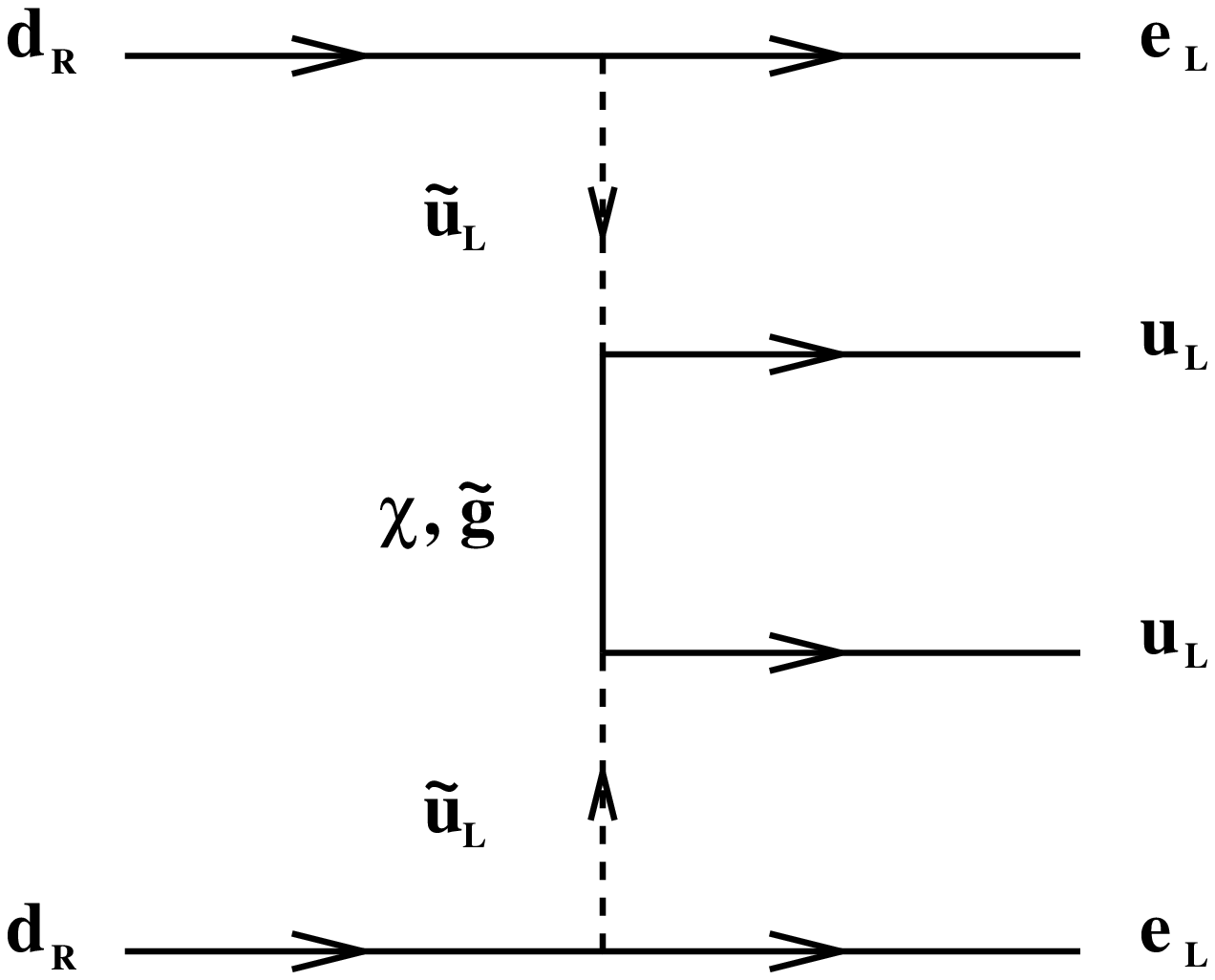}}
\put(70,0){\includegraphics{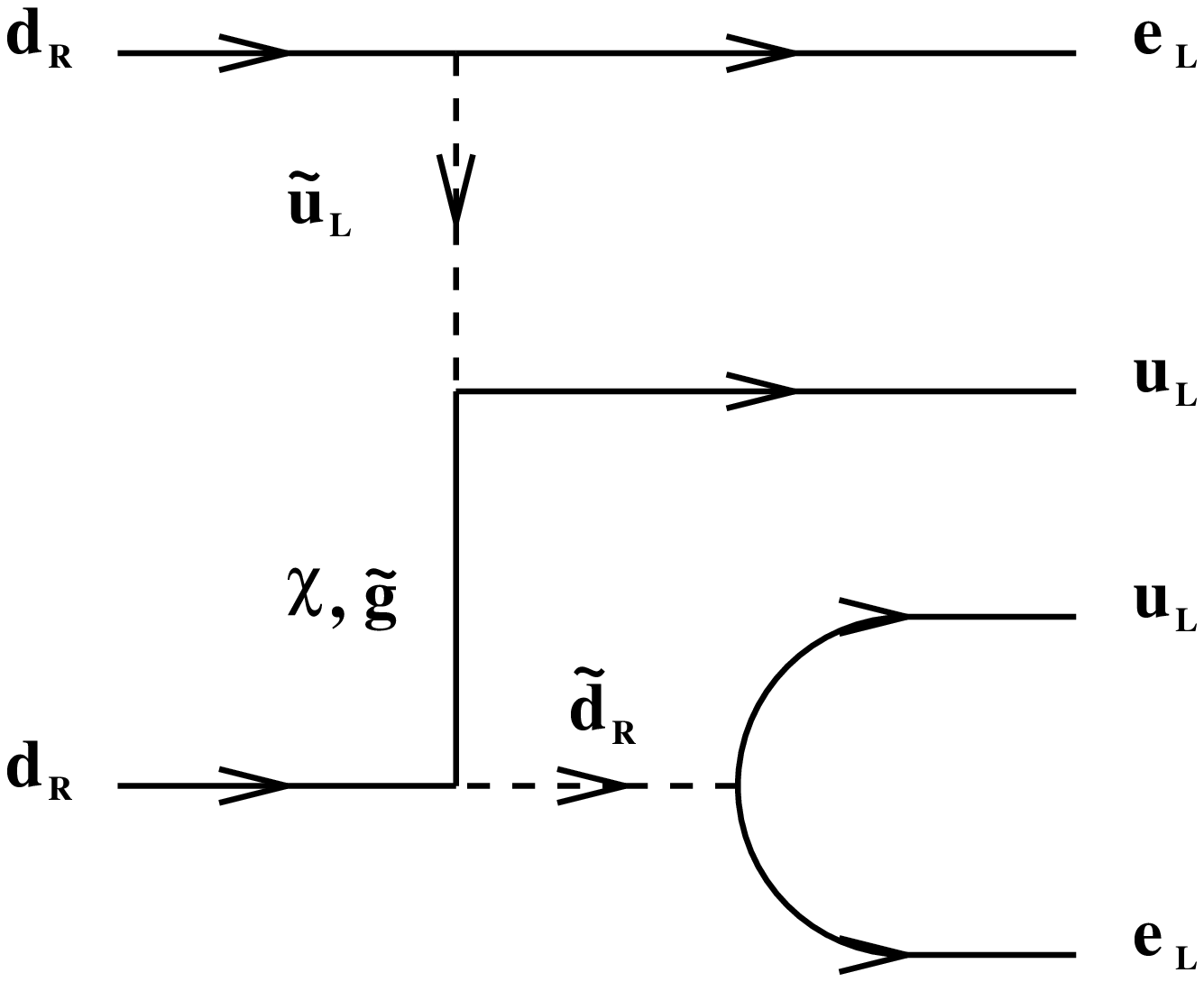}}
\end{picture}

\vspace{-2.cm}
\caption{Examples of Feynman graphs for $0\nu\beta\beta$ 
        decay within R--parity violating supersymmetric models
\protect\cite{KK-SprTracts00}.} 
\label{fig:R-Parit-Diagr}
\end{figure}

\begin{figure}[ht] 
\begin{picture}(100,70) 
\put(5,0){\includegraphics{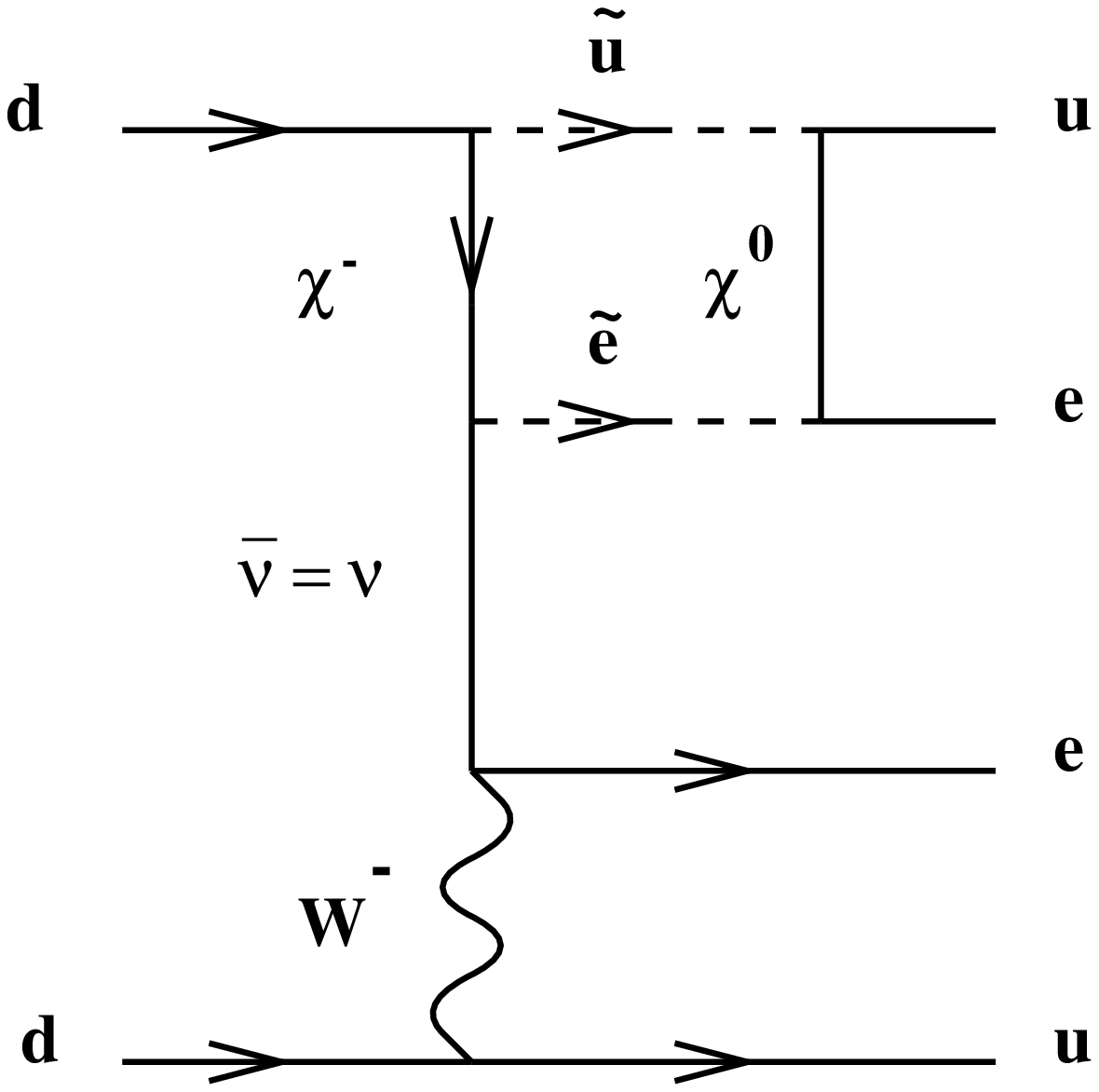}}
\put(70,0){\includegraphics{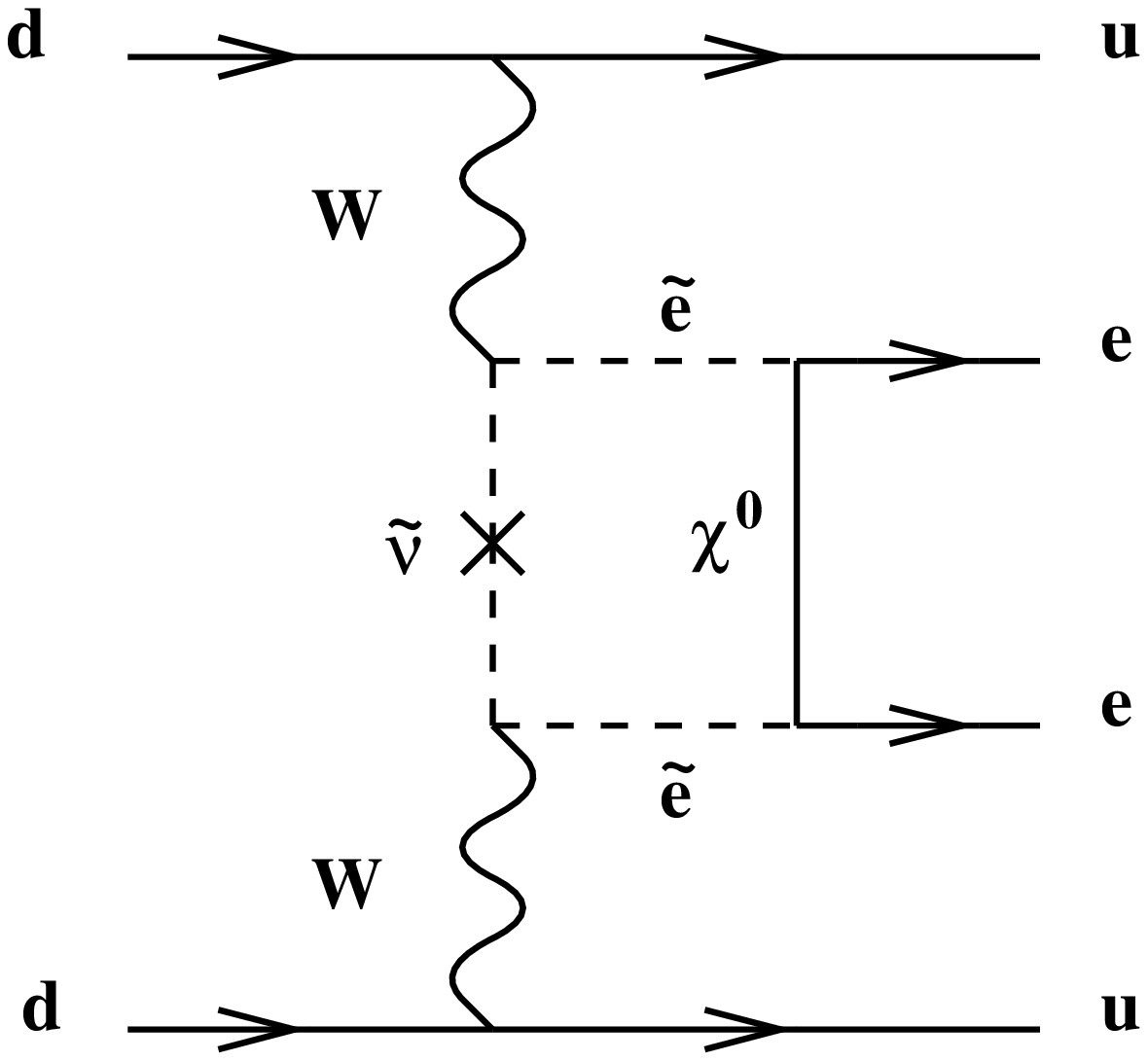}} 
\end{picture}

\vspace{-2.cm}
\caption{Examples of $R_P$ conserving SUSY contributions
        to $0\nu\beta\beta$ decay
\protect\cite{KK-SprTracts00}.} 
\label{fig:R_PSUSY-Diagr}
\end{figure}

\begin{figure}[ht] 
\begin{picture}(100,80) 
\put(5,0){\includegraphics{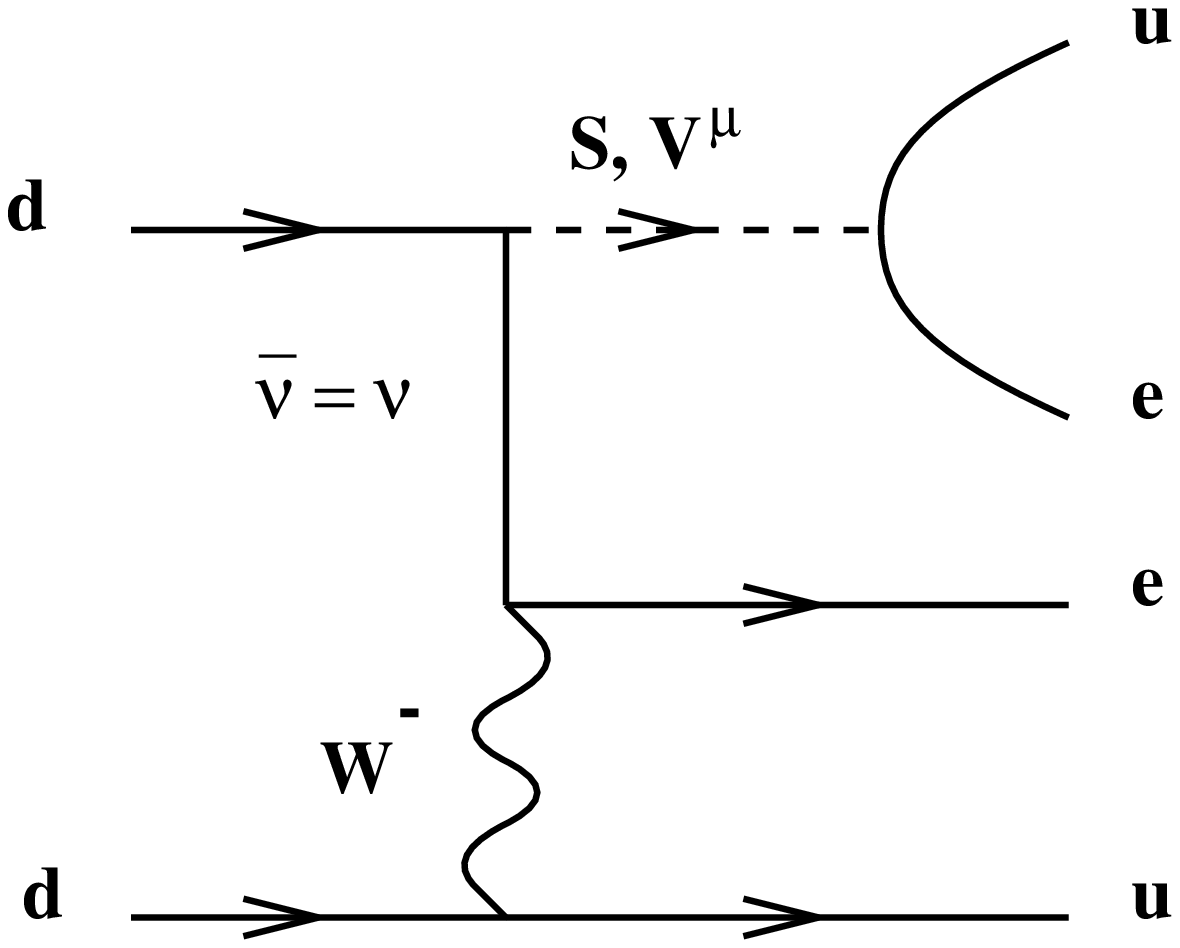}}
\put(70,0){\includegraphics{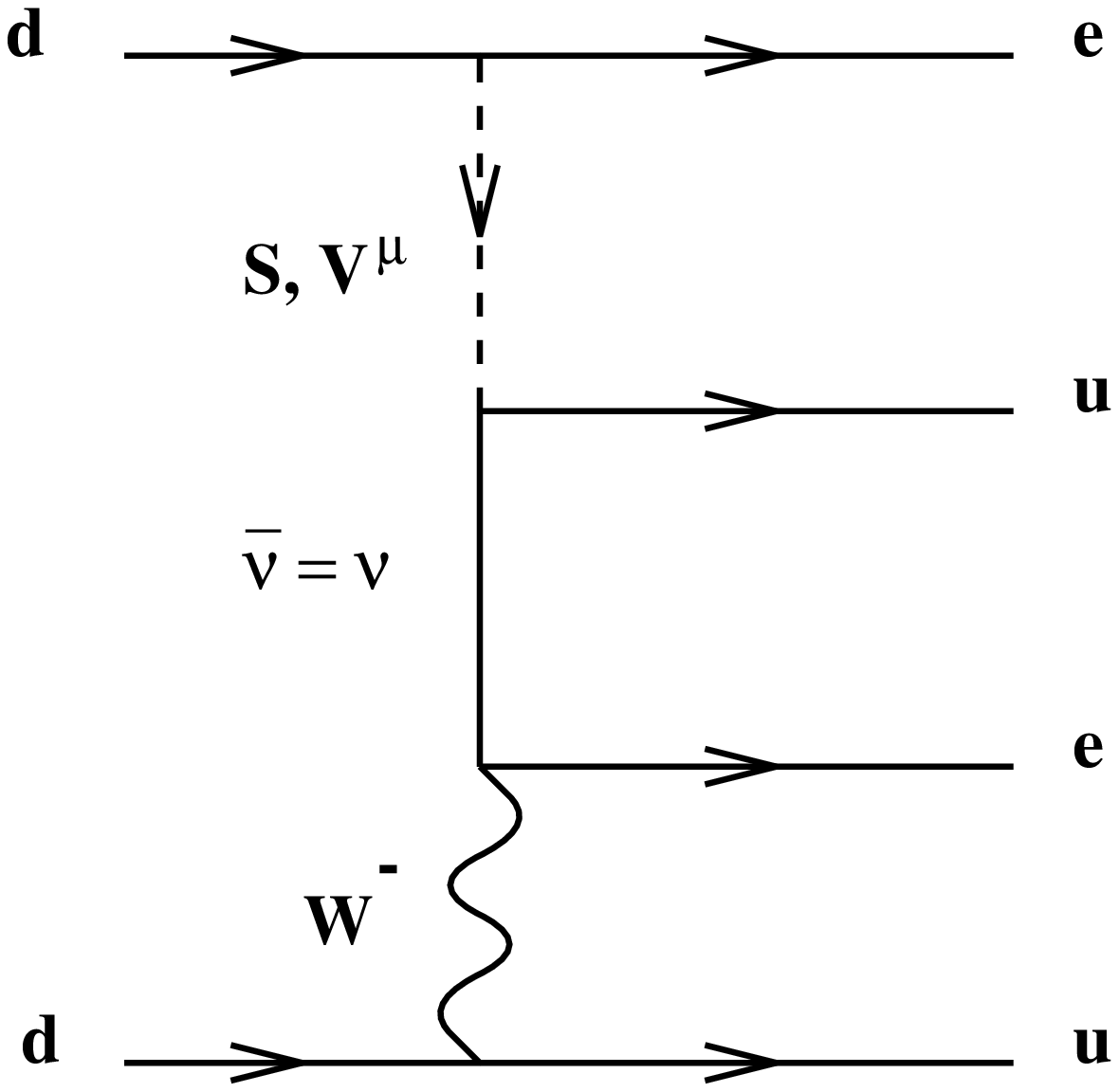}} 
\end{picture}

\vspace{-2.cm}
\caption{Examples of Feynman graphs for $0\nu\beta\beta$ decay 
        within LQ models. $S$ and $V^{\mu}$ stand for scalar 
        and vector LQs, respectively
\protect\cite{KK-SprTracts00}.} 
\label{fig:Scalar-Diagr}
\end{figure}

        {\bf Other beyond Standard Model Physics}:
        Assuming {\it other} mechanisms to dominate 
        the \znbb~ decay amplitude, 
        which have been studied extensively in our group, 
        and other groups, in recent years,  
        the result allows to set stringent limits on parameters of SUSY 
        models, leptoquarks, compositeness, masses of heavy neutrinos, 
        the right-handed W boson and possible violation of Lorentz 
        invariance and equivalence principle in the neutrino sector. 
        Figs.%
\ref{fig:R-Parit-Diagr},\ref{fig:R_PSUSY-Diagr},\ref{fig:Scalar-Diagr}, 
        show as examples some of the relevant graphs which 
        can in principle contribute to the \znbb amplitude 
        and from which bounds on the corresponding parameters 
        can be deduced assuming conservatively the measured half-life 
        as upper limit for the individual processes.

\begin{figure}[ht!] 
\begin{picture}(100,95) 
\put(15,110){\includegraphics{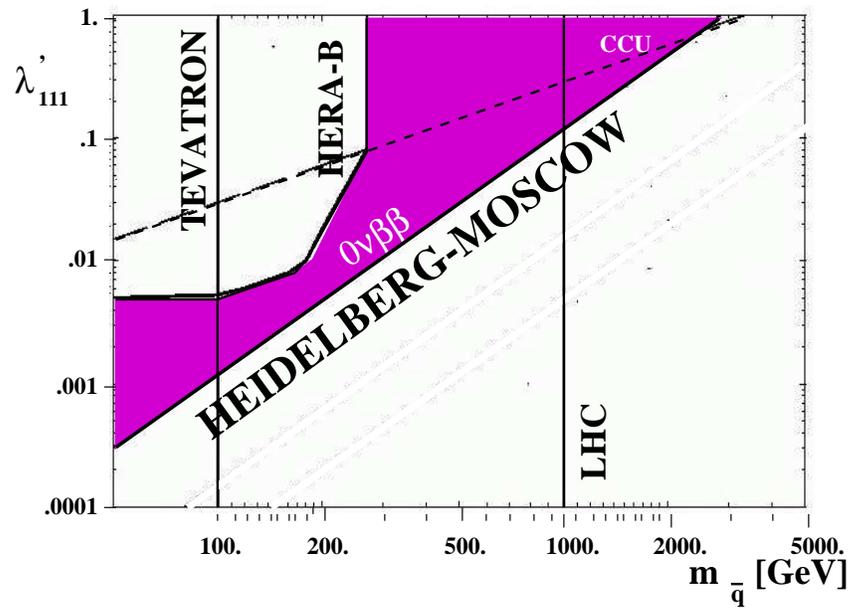}}
\end{picture}

\vspace{-2.cm}
\caption[]{Comparison of sensitivities of existing and future experiments on
        R$_p$-violating SUSY models in the plane 
        ${\lambda}^{'}_{111}$-$m_{\tilde{q}}$. 
        {\it Note the double logarithmic scale!}
        Shown are the areas currently excluded by the experiments 
        at the TEVATRON and HERA-B, 
        the limit from charged-current universality, 
        denoted by CCU, and the limit from  
        0$\nu\beta\beta$-decay from the \HM Experiment. 
        {\it The area beyond (or left of) the lines is excluded}. 
        The estimated sensitivity of LHC is also given (from 
\protect\cite{KK-SprTracts00}).} 
\label{fig:TEV-HERA-HM}
\end{figure}

\begin{figure}[ht] 
\begin{picture}(100,90) 
\put(15,100){\includegraphics{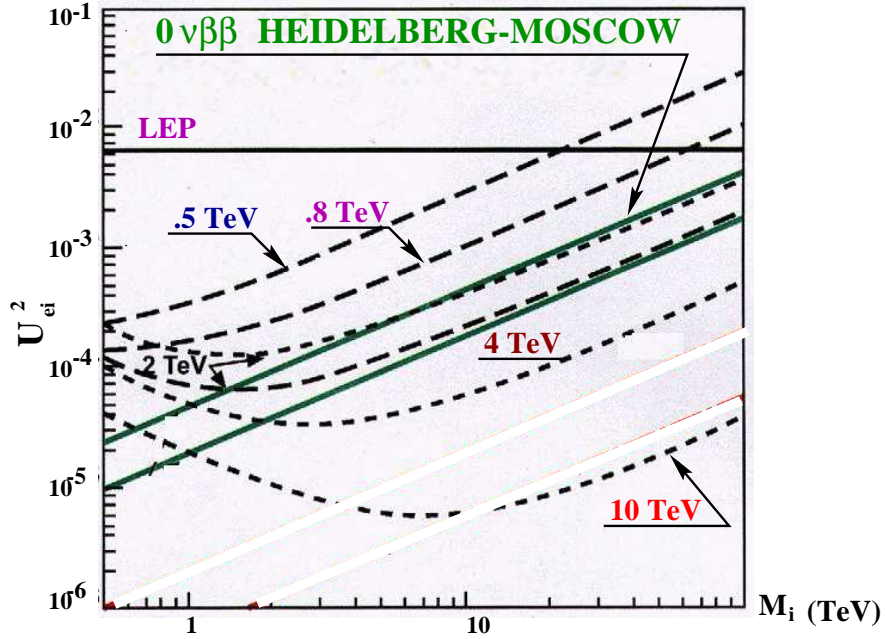}}
\end{picture}
\caption{Discovery limit for $e^- e^- \longrightarrow W^- W^-$ 
        at a linear collider as function of the mass M$_i$ of 
        a heavy left-handed neutrino, and of U$^2_{ei}$ for $\sqrt{s}$ 
        between 500 GeV and 10 TeV. In all cases the parameter space 
        above the line corresponds to observable events. 
        The limits from the \HM 0$\nu\beta\beta$ experiment 
        are shown also, the areas above the 0$\nu\beta\beta$ contour 
        line are excluded. The horizontal line denotes the limit 
        on neutrino mixing, U$^2_{ei}$, from LEP (from 
\protect\cite{Belan98}).
}
\label{fig:Belange}
\end{figure}

\begin{figure}[ht!] 
\begin{picture}(100,95) 
\put(10,110){\includegraphics{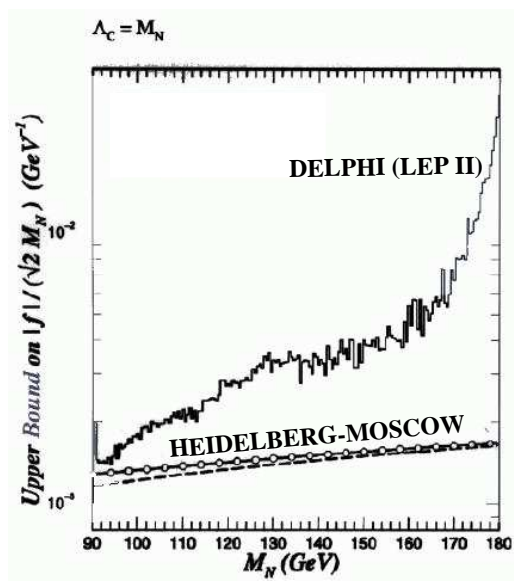}}
\end{picture}

\vspace{-2.cm}
\caption[]{Comparison between the $\beta\beta_{0\nu}$  \HM experiment 
        and the LEP II 
         upper bound on the quantity $|f|/(\sqrt{2}M_N)$ as a function of the
         heavy composite 
        neutrino mass $M_N$, with the choice $\Lambda_{\hbox{c}}=M_N$. 
         {\it Regions above the curves are excluded.} 
        The dashed and solid circle curves are 
        the $\beta\beta_{0\nu}$ bounds from the \HM experiment 
        (for details see 
\cite{Pan00}). 
}
\label{fig:Panella}
\end{figure}

\noindent
        Figs.%
\ref{fig:TEV-HERA-HM},\ref{fig:Belange} and \ref{fig:Panella} 
        show some results. 
        The most strict limit for the R-parity - breaking Yukawa 
        coupling ${\lambda}^{'}_{111}$ in R-violating SUSY models 
        is coming from 0$\nu\beta\beta$-decay. It is much stricter 
        then the limits obtained by accelerators, 
        whose limitation in energy is visible in Fig.
\ref{fig:TEV-HERA-HM} (from 
\cite{KK-SprTracts00}).

\newpage

\newpage
%
%

%

        The lower limit for super-heavy left-handed neutrino from the 
        0$\nu\beta\beta$ \HM experiment corresponds 
        to the discovery potential for the inverse 
        process $e^- e^- \longrightarrow W^- W^-$ of a linear 
        collider of 1-2 TeV (see Fig.%
\ref{fig:Belange}).
        The constraints concerning {\it composite} excited neutrinos 
        of mass M$-N$ obtained from 0$\nu\beta\beta$ decay (\HM 
	experiment) 
        are more strict than the results of LEPII, as shown in Fig.%
\ref{fig:Panella}.
        For a further discussion and for references we refer to 
\cite{KK60Y,KK-InJModPh98,KK-SprTracts00,KKS-INSA02}.


\section{Conclusion - Perspectives}

        Recent information from many {\it independent} sides seems 
        to condense now to a nonvanishing neutrino mass of the order 
        of the value found by the \HM experiment.
        This is the case for the results from 
        CMB, LSS, neutrino oscillations, particle theory and cosmology 
        (for a detailed discussion see 
\cite{KK-NewAn-PL04,KK-NewAn-NIM04,Bey03-BB}). 
        To mention a few examples: 
        Neutrino oscillations require in the case 
        of degenerate neutrinos common mass eigenvalues 
        of m $>$ 0.04\,eV.
        An analysis of CMB, large scale structure and X-ray 
        from clusters of galaxies yields a 'preferred' value 
        for $\sum m_\nu$ of 0.6\,eV 
\cite{Allen03-Wmap}. 
        WMAP yields $\sum m_\nu$ $<$ 1.0\,eV 
\cite{Hannes03},  
        SDSS yields $\sum m_\nu$ $<$ 1.7\,eV 
\cite{Teg03}.  
        Theoretical papers require degenerate neutrinos 
        with m $>$ 0.1, or 0.2\,eV or 0.3\,eV 
\cite{Moh03,BMV02,MaValle03,HirVall04,Ma-DARK02}, 
        and 
        the recent alternative cosmological concordance 
        model requires relic neutrinos with mass of order of eV 
\cite{S-Sarkar03}. 
        As mentioned already earlier 
\cite{HVKK-Bey02,KK-NewAn-NIM04} 
        the results of double beta decay and CMB measurements 
        together indicate that the neutrino mass eigenvalues 
        have the same CP parity, as required by the model of 
\cite{Moh03}.
        Also the approach of 
\cite{Hof04} 
        comes to the conclusion of a Majorana neutrino.
        The Z-burst scenario for ultra-high energy cosmic 
        rays requires $m_\nu$$\sim$ 0.4\,eV 
\cite{Farj00-04keV,FKR01}, 
        and also a non-standard model (g-2) has been connected 
        with degenerate neutrino masses $>$0.2\,eV 
\cite{MaRaid01}. 
        The neutrino mass determined from 0$\nu\beta\beta$ decay 
\       is consistent also with present models of leptogenesis 
        in the early Universe 
\cite{Reb03Bey}.
        It has been discussed that the Majorana nature of the neutrino 
        tells us that spacetime does realize a construct 
        that is central to construction of supersymmetric theories 
\cite{Ahl96}.

\begin{figure}[ht] 
\begin{picture}(100,190) 
\put(10,110){\includegraphics{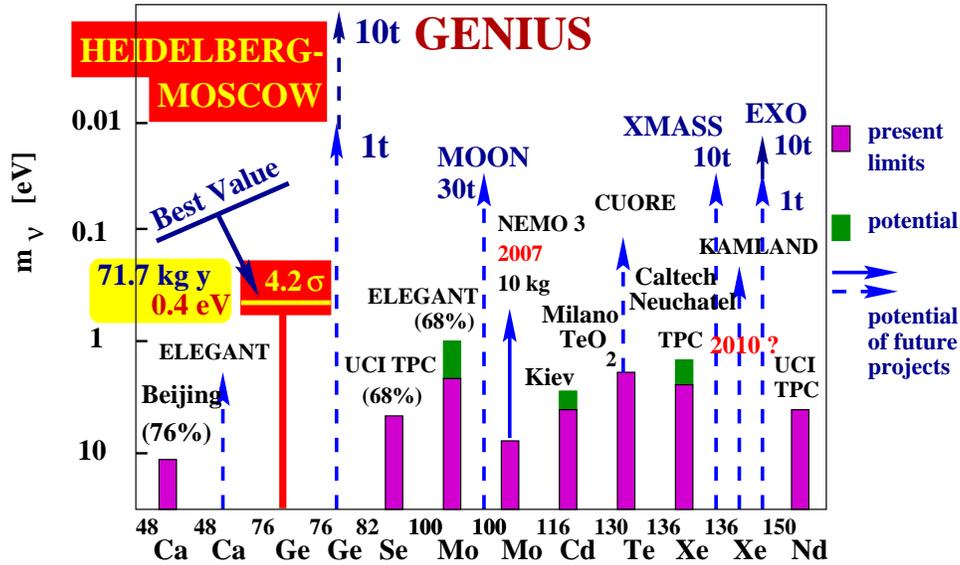}}
\end{picture}

\vspace{-10.cm}
\caption{
       Present sensitivity, and expectation for the future, 
       of the most promising $\beta\beta$ experiments. 
       Given are limits for $\langle m \rangle $, except 
        for the HEIDELBERG-MOSCOW experiment where 
        the measured {\it value} 
        is given (3$\sigma$ c.l. range and best value).
        Framed parts of the bars: present status; not framed parts: 
       future expectation for running experiments; solid and dashed lines: 
       experiments under construction or proposed, respectively. 
       For references see 
\protect\cite{KK60Y,KK02-PN,KK02-Found}.
\label{fig:Now4-gist-mass}}
\end{figure}

        {\sf Future:}
        With the \HM experiment, 
        {\it the era of small smart experiments is over.} 
        Fig. 
\ref{fig:Now4-gist-mass}
        shows the present result and a comparison 
        to the potential of the most sensitive other 
        double beta decay experiments and the possible 
        potential of some future projects.
        It is visible that the presently running experiments 
        have hardly a chance, to reach the sensitivity 
        of the \HM experiment.
        New approaches and considerably {\it enlarged} experiments would 
        be required to fix the \znbb half life with higher accuracy. 
        {\bf This will, however, only marginally improve the precision 
        of the deduced neutrino mass}, 
        because of the uncertainties in the nuclear matrix elements, 
        which probably hardly can be reduced to less than 50\%.  
        
        One has to keep in mind further, that {\bf no more can be learnt} 
        on {\it other}~ beyond standard model physics parameters from 
        future more sensitive experiments. The reason is that there 
        is {\bf a half-life} now, and {\bf no more a limit} on the half-life, 
        which could be further reduced.

        From future projects 
        one has to require that they should be able 
        to differentiate between a $\beta$ and a $\gamma$ signal, or 
        that the tracks of the emitted electrons should be measured. 
        At the same time, as is visible from the present information,  
        the energy resolution should be {\it at least} 
        in the order of that of Ge semiconductor detectors, 
        or better. 
        These requirements exclude at present calorimeter experiments 
        like CUORE, CUORICINO, 
        which {\it cannot} differentiate between a $\beta$ 
        and $\gamma$ signal, etc,  
        but also experiments like EXO 
\cite{EXO-LowNu2}, 
        {\it if} the latter will not be able to reconstruct 
        the tracks of the electrons, as it seems at present.  

	The {\it most discussed 'short-term' 
	"confirmation experiments"}  at present 
	are CUORICINO/CUORE and NEMO. Let us therefore, 
	to avoid usual misunderstandings, give a few comments.\\

{\bf CUORICINO, CUORE}:

	The general background problems of CUORICINO are illustrated 
	by the fact, that this experiment until now is not able to see 
	the \tnbb  decay of $^{130}{Te}$, whose half life is experimentally 
	known to be $T_{1/2}^{0\nu}$=(2.7$\pm$0.1)$\times$10$^{21}$ years 
\cite{Bern93}, 
	i.e. similar to the \tnbb halflife of $^{76}{Ge}$, which is 
 	very clearly seen in the \HM experiment 
	(see, e.g. 
\cite{KK-Doer03,HDM97}).
	The background in the range of Q$_{\beta\beta}$ 
	is for CUORICINO at present 
\cite{Broffer-Vened}
	a factor of two higher than in the \HM experiment.

	The present half-life limit for \onbb~ decay given 
	as 1.8$\times$10$^{24}$ years on a 90\% c.l. (statistical method 
	is not described, could however be important, see e.g. 
\cite{KK-PhysRC}),
	after a measuring time of 10.8\,kgy. The half-life 

	Corresponding to the effective neutrino mass deduced 
	from the \HM experiment, 
	for the case of $^{130}{Te}$ is according to 
\cite{Sta90}
	$T_{1/2}^{0\nu}$=2.5$\times$10$^{24}$\,y. 
	At a 90\% c.l. a corresponding limit could 
	be reached by CUORINO in additional 5\,months of continuous running, 
	i.e. realistically in more than a year. 
	Allowing an uncertainty in the calculated matrix element 
	of a factor of 2, however, could require a 16 times 
	larger measuring time, i.e. $\sim$30\,years, to make a statement 
	on a 90\% confidence level. 
	This means that {\it the CUORICINO experiment can, with good luck, 
	confirm the Heidelberg-Moscow result (on a reasonable   
	not only 90\% c.l.) in several years, but it can never disprove it}. 

	 The full version CUORE with about a factor of 15 larger 
	detector mass than CUORICINO  could, with background of CUORICINO, 
	have a sensitivity to probe the \HM result 
	in about one year of continuous measuring
 	on a 90\% confidence level. So unfortunately also this experiment 
	would require many years of measurement to make a statement 
	on a reasonable confidence level.\\ 

{\bf NEMO:}

        The NEMO project {\it can} see tracks, but unfortunately 
        has at present 
	only a small efficiency (14\%) , and {\bf a low energy resolution 
	of more than 200 keV}  not to talk about the background problems 
	from Rn.

	Therefore limits given for \onbb~ decay are lying at present 
\cite{Sarazin-Paris04}
	only at $T_{1/2}^{0\nu}$=1.9$\times$10$^{23}$\,y ($^{82}{Se}$) 
	and $T_{1/2}^{0\nu}$=3.5$\times$10$^{23}$\,y ($^{100}{Mo}$) 
	on a 90\% c.l, i.e. on a 1.5\,sigma level, for 0.55 and 5\,kg\,y 
	of measurement, respectively. To improve these limits 
	by a factor of 20, which is required (see 
\cite{Sta90})
	a t~~  l e a s t  , to check the results 
	of the \HM experiment presented in this paper, 
	the measurement times have to be increased by a factor of 400. 
	This means that this experiment is not able 
	to check the \HM result.\\

{\bf GENIUS:}

	An  i n  p r i n c i p l e  much more sensitive project 
	is probably the GENIUS project, proposed already in 1997 
\cite{KK-Bey97,KK-Hir-GENIUS,KK-Helm-Gen,KK-H-H-97,GEN-prop,KK-InJModPh98,KK60Y}.

        A GENIUS Test Facility, 
        (which could already be used to search for cold dark 
        matter by the annual modulation effect) 
        has started operation with 10\,kg 
        of natural Germanium detectors in liquid nitrogen
        in Gran-Sasso on May 5, 2003 
\cite{TF-Ra04,TF-NIM03,CERN03-GenTF,KK-Modul-NIM03},
	increased to 15 kg in October 2005. 
	The results from the GENIUS-Test-Facility show
\cite{Bi-KK03-NIM}, 
	however, 
	that though the search for cold dark matter should be feasible, 
	it may be technically rather 
	difficult, to increase the sensitivity of a 
	{\it GENIUS-like} experiment 
	for neutrinoless double beta decay beyond that 
	of the \HM experiment.

        However, if one wants 
        to get {\it independent} evidence for the neutrinoless 
        double beta  decay mode, one would probably, 
        wish to see the effect in {\it another} 
        isotope, which would then simultaneously give additional 
        information also on the nuclear matrix elements. 
        In view of these considerations, future efforts to obtain 
        {\it deeper} information on the process 
        of neutrinoless double beta decay, would require   
        {\it a new experimental approach, different from all, 
        what is at present persued}.

\vspace{0.3cm}
\noindent
{\bf Acknowledgement:}

\vspace{0.3cm}
\noindent
        The author would like to thank all colleagues, 
        who have contributed to the experiment over the last 15\,years. 
        He thanks in particular Irina Krivosheina, 
        for her important contribution to the analysis of this experiment. 

        Our thanks extend also to the technical staff of the 
        Max-Planck Institut f\"ur Kernphysik and 
        of the Gran Sasso Underground Laboratory.  
        We acknowledge the invaluable support from BMBF 
        and DFG, and LNGS of this project.
        We are grateful to the former State Committee of Atomic 
        Energy of the USSR for providing the enriched material 
        used in this experiment. 

	The author thanks Prof. M. Baldo Ceolin 
        for the kind invitation to give this talk.


\end{document}